\documentclass[english,amsmath,amssymb,superscriptaddress,prl,nofootinbib]{revtex4-1}
\usepackage[T1]{fontenc}
\usepackage[latin9]{inputenc}
\usepackage{geometry}
\usepackage{verbatim}
\usepackage{float}
\usepackage{units}
\usepackage{textcomp}
\usepackage{amsmath}
\usepackage{amssymb}
\usepackage{graphicx}
\usepackage{esint}

\makeatletter

\providecommand{\tabularnewline}{\\}

\@ifundefined{textcolor}{}
{%
 \definecolor{BLACK}{gray}{0}
 \definecolor{WHITE}{gray}{1}
 \definecolor{RED}{rgb}{1,0,0}
 \definecolor{GREEN}{rgb}{0,1,0}
 \definecolor{BLUE}{rgb}{0,0,1}
 \definecolor{CYAN}{cmyk}{1,0,0,0}
 \definecolor{MAGENTA}{cmyk}{0,1,0,0}
 \definecolor{YELLOW}{cmyk}{0,0,1,0}
 }

\usepackage{amsthm}
\usepackage{calrsfs}
\usepackage[T3,T1]{fontenc}
\usepackage{mathdesign}
\usepackage{textcomp}
\usepackage{fge}
\usepackage{MnSymbol}
\makeatletter 
\newcommand*{\shifttext}[2]{%
  \settowidth{\@tempdima}{#2}%
  \makebox[\@tempdima]{\hspace*{#1}#2}%
}
\makeatother
\DeclareMathAlphabet{\mathpzc}{OT1}{pzc}{m}{it}
\DeclareSymbolFont{tipa}{T3}{cmr}{m}{n}
\DeclareMathAccent{\invbreve}{\mathalpha}{tipa}{16}

\makeatother

\usepackage{babel}
\begin{document}

\title{An open source MATLAB program for fast numerical Feynman integral
calculations for open quantum system dynamics on GPUs}

\author{Nikesh S. Dattani}

\email{nike.dattani@chem.ox.ac.uk}

\affiliation{Physical and Theoretical Chemistry Laboratory, Department of Chemistry,
University of Oxford, Oxford, OX1 3QZ, UK}

\date{$\today$}
\begin{abstract}
This MATLAB program calculates the dynamics of the reduced density
matrix of an open quantum system modeled by the Feynman-Vernon model.
The user gives the program a vector describing the coordinate of an
open quantum system, a hamiltonian matrix describing its energy, and
a spectral distribution function and temperature describing the environment's
influence on it, in addition to the open quantum system's intial density
matrix and a grid of times. With this, the program returns the reduced
density matrix of the open quantum system at all (or some) moments
specified by that grid of times. This overall calculation can be divided
into two stages: the setup of the Feynman integral, and the actual
calculation of the Feynman integral for time-propagation of the density
matrix. When this program calculates this propagation on a multi-core
CPU, it is this propagation that is usually the rate limiting step
of the calculation, but when it is calculated on a GPU, the propagation
is calculated so quickly that the setup of the Feynman integal actually
becomes the rate limiting step for most cases tested so far. The overhead
of transfrring information from the CPU to the GPU and back seems
to have negligible effect on the overall runtime of the program. When
the required information cannot fit on the GPU, the user can choose
to run the entire program on a CPU.
\end{abstract}
\maketitle

\section{Introduction\label{sec:introduction}}

One is very often interested in how the reduced density matrix $\rho$
of an \textbf{OQS} (\textbf{o}pen \textbf{q}uantum \textbf{s}ystem)
changes with respect to time ($\rho D\equiv$reduced density matrix
dynamics). The most popular open quantum system model to date is the
Feynman-Vernon model (see section II %
\ref{sec:setting}), and for this model, a formally exact mathematical
description of the $\rho D$ was deveoped by Richard P. Feynman and
two of his PhD students Frank L. Vernon Jr. and Willard H. Wells in
the late 1950s \cite{1961Wells,1963Feynman,RichardPhillipsFeynmanAlbertR.Hibbs2010}.
In this mathematical formalism, the $\rho D$ is represented in terms
of Feynman integrals, for which no general closed form analytic solution
is known. Therefore, to determine the numerical values of the elements
of the reduced density matrix at a given point in time, one needs
to evaluate these Feynman integrals numerically. Early work in numerically
calculating these Feynman integrals used Monte Carlo methods, but
since the integrnds are highly oscillatory, deterministic algorithms
have been the most popular methods for these numercal Feynman integral
calculations since breakthroughs in algorithmic development were made
, mainly between 1992 and 2001. 

The \textbf{qu}asi-\textbf{a}diabatic \textbf{p}ropagator Feynman%
\footnote{The term `path integral' is used more commonly than `Feynamn integral'
here, but this term is ambiguous. Currently, the first result on the
search engine at www.google.com, when the search query `path integral'
is entered, is a Wikipedia page that currently links to three different
meanings of the word `path integral': (1) line integral, (2) functional
integration, and (3) path integral formulation. Only the third of
these is unambiguously the Feynman integral discussed in this paper.
The `line integral' is an integral over a path, rather than over a
set of paths; and the term `functional integral' can refer to at least
three types of functional integrals: (1) the Wiener integral, (2)
the Lévy integral, and (3) the Feynman integral.%
} \textbf{i}ntegral (\textbf{QUAPI}) technique indroduced by Nancy
Makri in 1992\cite{1992Makri} helps the numerical calculation of
these Feynman integrals converge with a larger sized time step used
in the discretization of the Feynman integrals, when the bath is nearly
adiabatic. Today's most popular algorithm for calculating these numerical
Feynman integrals (whether using a quasi-adiabatic propagotor or not)
uses a representation of the numerical Feynman integrals in terms
of matrix-vector-like operations, and is called the `tensor propagator
scheme'. The most evolved form of the tensor propagator scheme was
explained by Nancy Makri and Dmitrii E. Makarov in 1995 \cite{1995Makri2},
although the original formulation of it was introduced by them in
1994\cite{1994Makarov} and explained in more detail in their 1995
paper\cite{1995Makri}. The \textbf{f}iltered \textbf{p}ropagator
\textbf{f}unctional (\textbf{FPF)} introduced by Eunji Sim and Nancy
Makri in 1996\cite{1996Sim,Sim1997a} is an improvement of the tensor
propagator scheme which saves computer memory by using Monte Carlo
importance sampling to filter which Feynman paths are included in
the Feynman integral, but Eunji Sim's 2001 algorithm \cite{2001Sim}
is an alternative way of filtering the Feynman paths within the tensor
propagator scheme which does not require Monte Carlo importance sampling
and has been shown to save even more memory than the FPF technique,
without sacrificing accuracy. I am not aware of any further algorithmic
developments for the calculations of numerical Feynman integrals for
the $\rho D$ of the Feynman-Vernon model in the first decade of the
21st century. Very recently another technique was introduced which
can significantly reduce the number of Feynman paths required in the
calculation of these Feynman integrals without significantly deteriorating
the accuracy, which is particularly useful when the reduced density
matrix changes so quickly that resolving its dynamics requires it
to be calculated much more often than the response function of the
OQS's environment needs to be sampled for a chosen amount of accuracy\cite{Dattani2012}. 

\medskip{}

In the MATLAB program described in this paper, the main calculation
which propagates the reduced density matrix in time using Feynman
integration, is mainly carried out by the MATAB function BSXFUN (\textbf{b}inary
\textbf{s}ingleton e\textbf{x}pansion \textbf{fun}ction), which was
first introduced in March 2007 in version R2007a of MATLAB. This algorithm
which uses BSXFUN is the fastest MATLAB implementation for the tensor
propagator scheme metioned above, with which I have experimented.
It is also extremely well suited for performance on \textbf{GPU}s
(\textbf{g}raphical \textbf{p}rocessing \textbf{u}nits). 

The company AccelerEyes developed the commercial software Jacket (for
which the 1st beta version was v0.2 which was released in June 2008),
which allows MATLAB functions to be performed with significant speedup
on GPUs. BSXFUN was first incorporated into Jacket in v1.2, which
was released in October 2009. In April 2010 a viral blog entry was
published\cite{2010Pryor}, entitled `Crushing MATLAB Loop Runtimes
with BSXFUN', which demonstrated that for the example calculation
in that study, BSXFUN performed about 45 times faster than the naive
MATLAB algorithm, when both calculations were executed on the same
CPU. Even more impressively, when this BSXFUN calculaton was performed
on their GPU, the calculation was sped up by almost another 6 times!

MATLAB itself did not have a built-in capability to execute caculations
on GPUs until September 2010 in version R2010b, and it did not support
the use of the built-in BSXFUN function on GPUs until very recently
(March 2012 in version R2012a).

\section{Setting\label{sec:setting} }

The most popular OQS model is currently the Feynman-Vernon model.
In the Feynman-Vernon model, the OQS (whose coordinate is denoted
by $s$) is coupled linearly to a set of \textbf{QHO}s (\textbf{q}uantum
\textbf{h}armonic \textbf{o}scillators) $Q_{k}$:
\begin{alignat}{1}
H & =H_{{\rm OQS}}+H_{{\rm OQS-bath}}+H_{{\rm bath}}\\
 & =H_{{\rm OQS}}+\sum_{\kappa}c_{\kappa}sQ+\sum_{\kappa}\big(\textrm{\textonehalf}m_{\kappa}\dot{Q_{\kappa}}^{2}+\text{\textonehalf}m_{\kappa}\omega_{\kappa}^{2}Q_{\kappa}^{2}\big)\,.\label{eq:hamiltonianFV}
\end{alignat}
In most models the QHOs span a continuous spectrum of frequencies
$\omega_{\kappa}$ and the strength of the coupling between the QHO
of frequency $\omega$ and the OQS is given by the spectral distribution
function $J(\omega)$:

\begin{equation}
J(\omega)=\frac{\pi}{2}\sum_{\kappa}\frac{c_{\kappa}^{2}}{m_{\kappa}\omega_{\kappa}}\delta(\omega-\omega_{\kappa})\,.\label{eq:spectralDistribution}
\end{equation}
For the hamiltonian of the Feynman-Vernon model, the bath response
function $\alpha(t)$ is the following integral transform of $J(\omega)$,
in terms of the inverse temperature $\beta\equiv\frac{1}{k_{B}T}$:

\begin{align}
\alpha(t) & =\frac{1}{\pi}\int_{0}^{\infty}J(\omega)\Big(\coth\Big(\frac{\beta\omega\hbar}{2}\Big)\cos(\omega t)-{\rm i}\sin(\omega t)\Big){\rm d}\omega\label{eq:bathResponseFunction}\\
 & =\frac{1}{\pi}\int_{-\infty}^{\infty}\frac{J(\omega)\exp\Big(\frac{\beta\omega\hbar}{2}\Big)}{2\sinh\Big(\frac{\beta\omega\hbar}{2}\Big)}e^{-{\rm i}\omega t}{\rm {\rm d}}\omega,\, J(-\omega)\equiv J(\omega)\label{eq:bathResponseFunctionAsMakri'sFourierTransform}\\
 & =\frac{1}{\pi}\int_{-\infty}^{\infty}\frac{J(\omega)}{1-\exp(-\beta\omega\hbar)}e^{-{\rm i}\omega t}{\rm d}\omega\,,\, J(-\omega)\equiv J(\omega),\label{eq:bathResponseFunctionAsFourierTransformOfBoseFunction}
\end{align}
where, equation \ref{eq:bathResponseFunctionAsFourierTransformOfBoseFunction}
can be written in terms of the Bose-Einstein distribution function
with $x=\beta\omega\hbar$:

\begin{equation}
f^{{\rm Bose-Einstein}}(x)=\frac{1}{1-\exp(-x)}.\label{eq:boseEinsteinFunction}
\end{equation}

Assuming that the density matrix of the OQS plus its environment at
$t=0$ is $\rho_{{\rm total}}=\rho(0)\otimes e^{-\beta H_{{\rm bath}}}$,
the reduced density matrix elements at time $t=N\cdot\Delta t$ are
given exactly by\cite{1995Makri3} (the integrals are used if the
coordinate $s$ of the OQS is continuous, and the summations are used
if it is discrete or discretized):

\begin{equation}
\langle s_{N}^{+}|\rho(t)|s_{N}^{-}\rangle=\sumint\lim_{\substack{\Delta t\to0\\
\Delta k_{{\rm max}}\to\infty
}
}\left(\prod_{k=0}^{N-1}\langle s_{k+1}^{+}|e^{-\frac{{\rm i}}{\hbar}H_{{\rm OQS}}\Delta t}|s_{k}^{+}\rangle\langle s_{k}^{-}|e^{\frac{{\rm i}}{\hbar}H_{{\rm OQS}}\Delta t}|s_{k+1}^{-}\rangle\right)\langle s_{0}^{+}|\rho(0)|s_{0}^{-}\rangle I\left(\{s_{k}^{\pm}\}_{k=0}^{N};\Delta t\right)\prod_{k=0}^{N-1}{\rm d}s_{k}^{+}{\rm d}s_{k}^{-}\,\,,
\end{equation}

where the discretized influence functional%
\footnote{which is actually a function%
} can be written as%
:

\begin{equation}
I=\exp\left(-\sum_{k=0}^{\Delta k_{{\rm max}}}\sum_{k^{\prime}=0}^{k}(s_{k}^{+}-s_{k}^{-})(\eta_{kk^{\prime}}s_{k^{\prime}}^{+}-\eta_{kk^{\prime}}^{*}s_{k^{\prime}}^{-})\right)\,\,,
\end{equation}

and the $\eta$ coefficients are given in terms of the bath response
function $\alpha(t)$ by \cite{Dattani2012b}:

\begin{alignat}{1}
\eta_{kk^{\prime}} & \equiv\int_{k\Delta t}^{(k+1)\Delta t}\hspace{-1.85mm}\int_{k^{\prime}\Delta t}^{(k^{\prime}+1)\Delta t}\alpha(t^{\prime}-t^{\prime\prime}){\rm d}t^{\prime\prime}{\rm d}t^{\prime}\,\,,\, k\ne k^{\prime}\,\,,\label{eq:etaKK'generalTrotter}\\
\eta_{kk} & \equiv\int_{k\Delta t}^{(k+1)\Delta t}\hspace{-1.85mm}\int_{k\Delta t}^{t^{\prime}}\alpha(t^{\prime}-t^{\prime\prime}){\rm d}t^{\prime\prime}{\rm d}t^{\prime}\,\,,\, k\notin\{0,N\}\,\,,\label{eq:etaKKgeneralTrotter}\\
\eta_{N0} & \equiv\int_{N\Delta t-\nicefrac{\Delta t}{2}}^{N\Delta t}\quad\int_{0}^{\nicefrac{\Delta t}{2}}\alpha(t^{\prime}-t^{\prime\prime}){\rm d}t^{\prime\prime}{\rm d}t^{\prime}\,\,,\label{eq:etaN0generalStrang}\\
\eta_{00} & \equiv\int_{0}^{t}\hspace{-1.85mm}\int_{0}^{t^{\prime}}\alpha(t^{\prime}-t^{\prime\prime}){\rm d}t^{\prime\prime}{\rm d}t^{\prime}\,\,,\label{eq:eta00generalStrang}\\
\eta_{NN} & \equiv\int_{N\Delta t-\nicefrac{\Delta t}{2}}^{N\Delta t}\quad\int_{N\Delta t-\nicefrac{\Delta t}{2}}^{t^{\prime}}\alpha(t^{\prime}-t^{\prime\prime}){\rm d}t^{\prime\prime}{\rm d}t^{\prime}\,\,,\label{eq:etaNNgeneralStrang}\\
\eta_{k0} & \equiv\int_{k\Delta t}^{(k+1)\Delta t}\hspace{-1.85mm}\int_{0}^{\nicefrac{\Delta t}{2}}\alpha(t^{\prime}-t^{\prime\prime}){\rm d}t^{\prime\prime}{\rm d}t^{\prime}\,\,,\label{eq:etaK0generalStrang}\\
\eta_{Nk} & \equiv\int_{N\Delta t-\nicefrac{\Delta t}{2}}^{N\Delta t}\quad\int_{k\Delta t}^{(k+1)\Delta t}\alpha(t^{\prime}-t^{\prime\prime}){\rm d}t^{\prime\prime}{\rm d}t^{\prime}\,\,,\label{eq:etaNkgeneralStrang}
\end{alignat}
and can very often be represented by closed form analytic expressions\cite{Dattani2012b,Dattani2012c}.
The purpose of the program is to calculate all (or some) of the matrix
elements $\langle s_{N}^{+}|\rho(t)|s_{N}^{-}\rangle$ at a time $t$
(or at all of the times $\{t_{k}=k\cdot\Delta t\}_{k=0}^{N}$), given
the system coordinate $s$, its hamiltonian $H_{{\rm OQS}}$ and initial
density matrix $\rho(0)$, the spectral distrution function $J(\omega)$
\textbf{or} the bath response function $\alpha(t)$, the temperature
$T$, and the convergence parameters $\{\Delta t,\Delta k_{{\rm max}}\}$.

To save time in recalucalting certain quantities involved in the overall
calculation, the program stores four variables when $N>\Delta k_{\max}$,
each containing 
\begin{equation}
M^{2(\Delta k_{{\rm max}}+1)}\label{eq:memoryComplexity}
\end{equation}
(double precision) floating point complex numbers (here $M$ is the
dimension of the Hilbert space of the OQS). In my experience, using
single precision numbers has sometimes given values for $\rho(t)$
that are very slightly (but observably) different from when double
precision numbers are used, and since the purpose of the numerical
Feynman integral technique described here is to calculate \textit{exact
}values of $\rho(t)$ (usually for benchmarking less accurate, less
computationally expensive methods), double precision is recommended
(although the program will still work if one desires to keep less
digits to save memory). When these complex numbers are double precision,
they take up 16 bytes of the computer's memory, so the overall amount
of memory required to store the four variables mentioned just before
equation \ref{eq:memoryComplexity} is

\begin{alignat}{1}
{\rm PMC} & =4\cdot16\cdot M^{2(\Delta k_{{\rm max}}+1)}\\
 & =64\cdot M^{2(\Delta k_{{\rm max}}+1)}\label{eq:primaryMemoryCost}
\end{alignat}
bytes of information overall (here \textbf{PMC} stands for \textbf{p}rimay
\textbf{m}emory \textbf{c}ost, %
{} since the storage of these numbers almost always constitutes the
vast majority of the memory required for the overall calculation,
and the PMC is the most important computational complexity measure
associated with the program described in this paper). This memory
cost could be reduced if one of the filtering techniques%
{} mentioned in the introduction section of this paper were to be implemented;
however the current version of the program does not have either of
these algorithms implemented.%

\section{Model system for example calculations\label{sec:modelSystemForExampleCalculations}}

All example calculations in this paper will be for a model system
with:

\begin{alignat}{1}
s & \in\{0,1\}\,\,,\\
H_{{\rm OQS}} & =\frac{1}{2}\begin{pmatrix}0 & \Omega(t)\\
\Omega(t) & 0
\end{pmatrix}\,\,,\,\textrm{and}\label{eq:H_OQS}\\
J(\omega) & =A\omega^{3}e^{(-\nicefrac{\omega}{{\rm \omega_{c}}})^{2}\,\,.}\label{eq:superOhmicJwithGaussianCutoff}
\end{alignat}
This OQS model has been used to describe many experiments (some examples
are \cite{Ramsay2010}%
) on laser-driven single quantum dots after a \textbf{r}otating \textbf{w}ave
\textbf{a}pproximation (\textbf{RWA}). The specific parameters used
in this paper will consistently be:

\begin{alignat}{1}
\Omega(t) & =\nicefrac{\pi}{8}\,\,,\nonumber \\
A & =\pi0.027{\rm ps^{2}}\,\,,\label{eq:AforJ(w)usedInThisPaper(fromRamsayGoddenEtc2010prl)}\\
\omega_{c} & =2.2{\rm ps^{-1}}\,\,,\,\text{and}\label{eq:wcForJ(w)usedInThisPaper(fromRamsayGoddenEtc2010prl)}\\
T & =25{\rm K}\,\,.\nonumber 
\end{alignat}
The values of $A$ and $\omega_{c}$ come from the experimental study
of this model system described in \cite{Ramsay2010}, and have been
used in many computational studies that use this same model system
(some examples are \cite{McCutcheon2010,McCutcheon2011a,Dattani2012}).
The values of $\Omega$ and $T$ were chosen in order for the damped
Rabi oscillations to last for a long time (for better evaluation of
the performance of the program's propagation of $\rho$ with respect
to time), while being damped enough to be physically interesting,
and while keeping $\Omega$ and $T$ physically realizable%
.

\section{Performance\label{sec:performance}}

All calculations reported in this paper were performed on an Intel
Xeon/Nehalem CPU with 2.8GHz clock rate, and/or on an Nvidia C2020
GPU which are part of the SKYNET Viglen/NVIDIA hybrid GPGPU Cluster
provided by the Oxford Supercomputing Centre at the Oxford e-Research
Centre at University of Oxford. This this cluster is continually in
use by many people, making its performance variant with respect to
time. This may be reflected in some of the performance times reported
below, but it is still valuable to see the overall trends and comparisons.

\begin{center}
\begin{table}[H]
\caption{Running the main propagation calculations ($k>\Delta k)$) on the
CPU vs the GPU for varying values of $\Delta k_{{\rm max}}$. The
pimary memory cost (PMC) associated with each value of $\Delta k_{{\rm max}}$
is given by equation \ref{eq:primaryMemoryCost}.\label{tab:comparisonWithVaryingDeltaKmax}}

\centering{}%
\begin{tabular*}{0.95\textwidth}{@{\extracolsep{\fill}}ccccc}
$\Delta k_{{\rm max}}$ & PMC & Run time on CPU (seconds) & Run time on GPU (seconds) & Approximate speedup on GPU\tabularnewline
\hline 
\noalign{\vskip\doublerulesep}
\hline 
2 & 4.096 KB & 0.071580 & 1.224592 & \tabularnewline
3 & 16.38 KB & 0.081609 & 1.226297 & \tabularnewline
4 & 65.54 KB & 0.107413 & 1.230001 & \tabularnewline
5 & 262.1 KB & 0.209927 & 1.22675 & \tabularnewline
6 & 1.049 MB & 0.641943 & 2.177962 & \tabularnewline
7 & 4.194 MB & 2.571893 & 1.286302 & 2x\tabularnewline
8 & 16.78 MB & 16.347692 & 1.766242 & 9x\tabularnewline
9 & 67.11 MB & 59.452344 & 4.673941 & 13x\tabularnewline
10 & 268.4 MB & 291.731064 & 15.580798 & 19x\tabularnewline
11 & 1.074 GB & 1198.620937 & 21.130031 & 18x\tabularnewline
12 & 4.295 GB & 4911.546824 & \multicolumn{2}{c}{Required data did not fit on the GPU}\tabularnewline
\end{tabular*}
\end{table}

\par\end{center}

Based on the results in table \ref{tab:comparisonWithVaryingDeltaKmax},
we see that the speed benefit gained by running the main propagation
calculations on the GPU rather than the CPU become more and more apparent
as $\Delta k_{{\rm max}}$ is increased, until $\Delta k_{{\rm max}}=11$.
For this reason, the performance tests reported below are for calculations
where $\Delta k_{{\rm max}}=11$.

\begin{figure}[H]
\begin{centering}
\caption{Top: Profiler output when the main propagation calculations are performed
on the CPU (top) and GPU (bottom). \label{fig:profilerForCPUvsGPU}}

\par\end{centering}

\begin{centering}
\includegraphics[width=0.9\textwidth]{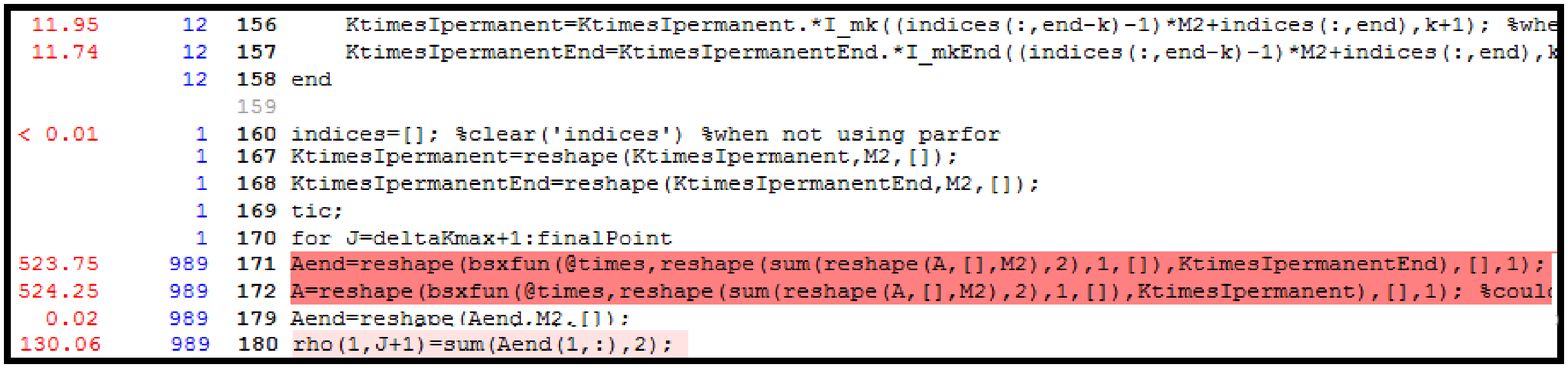}
\par\end{centering}

\begin{centering}
\includegraphics[width=0.9\textwidth]{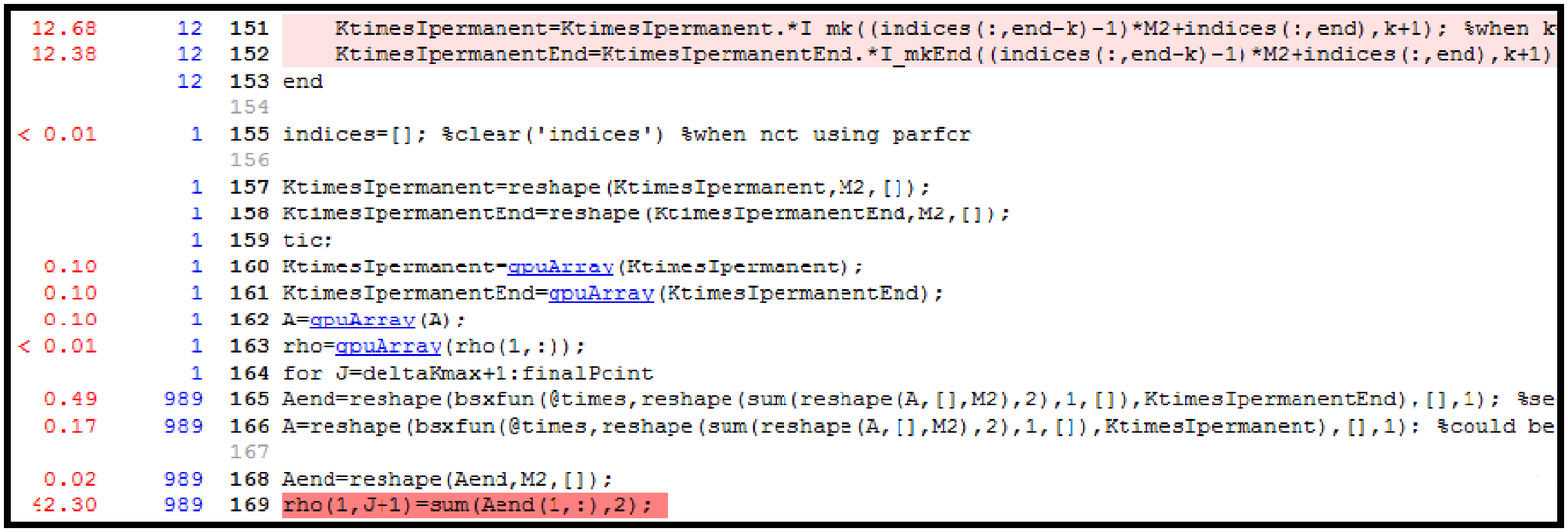}
\par\end{centering}

\end{figure}

Figure \ref{fig:profilerForCPUvsGPU} presents an estimate of the
amount of time spent by the program on each line of the part of the
code surrounding the lines that execute the main propagation calculations
(lines 170-171 for the CPU version of the program, and lines 164-165
for the GPU version). The calculation of the summation to determine
$\rho$ at the times $\{t_{k}=k\cdot\Delta t\}_{k=\Delta k_{{\rm max}}}^{N}$%
{} is executed on line 180 for the CPU version and 169 for the GPU version.

The most striking observation from figure \ref{fig:profilerForCPUvsGPU}
is that while on the CPU version, the propagation calculations (lines
170-171) are the obvious bottleneck, when these calculations are done
on the GPU (lines 165-166 of the GPU version), the profiler estimates
that these calculations take less than a second in total (despite
being run 989 times each) ! This means that the new bottleneck for
the GPU version is the summation on line 169, that caculates $\rho$
at 989 successive points in time, and according to the profiler's
estimate takes 42.3 seconds in total (0.0428 seconds for each time
value at which $\rho$ is calculated). 

Very often, one is only interested in $\rho(t)$ after a certain amount
of time $t_{{\rm specific}}$ has passed. For example, in the study
of laser-driven quntum dots, one is often interested in the population
{} of excitons after a certain amount of time (given by $\langle1|\rho(t_{{\rm specific}})|1\rangle=\rho_{11,t_{{\rm specific}}}$
in the model studied in this paper), and how this quantity changes
with respect to a physical parameter. The experimental data in figure
1 of \cite{2010Ramsay} and in figures 1 and 2 of \cite{Ramsay2010}
report the photocurrent at a specific time $t_{{\rm specific}}$ (the
photocurrent is related to $\rho_{11,t_{{\rm specific}}}$), as a
function of the `pulse area' (which is related to $\Omega(t)$ in
our hamiltonian given by equation \ref{eq:H_OQS}) of the laser that
drives a quantum dot. Various figures in theoretical and computatonal
studies of this same system (see for example \cite{2007Vagov,McCutcheon2010,2011Vagov,McCutcheon2011a,Glassl2011a,Luker2012},
and many more) present $\rho_{11,t_{{\rm specific}}}$ as a function
of the pulse area. In all of these cases, one is not interested in
$\rho(t)$ at every point in time, but only in $\rho(t_{{\rm specific}})$,
so line 169 of the GPU version (or line 180 of the CPU vesion) of
the code in figure \ref{fig:profilerForCPUvsGPU} would only have
to be executed once instead of 989 times, for each pulse area for
which one desires to know $\rho(t_{{\rm specific}})$. 

\begin{figure}
\caption{CPU time and GPU time vs $t_{{\rm specific}}$ when $\rho$ is only
calculated at the one time $t_{0}\equiv t_{{\rm specific}}$.\label{fig:cpuTimeAndGPUtimeAsFunctionOfT_specific}}

\includegraphics[width=0.6\textwidth]{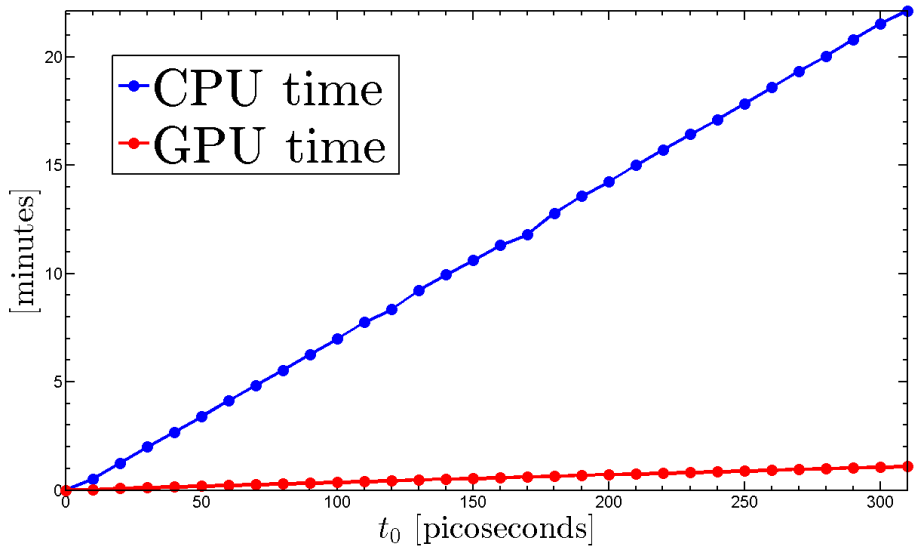}
\end{figure}

For this reason, the program described in this paper allows the user
to specify the input variable \texttt{allPointsOrJustFinalPoint} by
either the string \texttt{`allPoints'}, which runs line 180 of the
CPU version, or line 169 of the GPU version, for all values of $\{t=k\cdot\Delta t\}_{k=\Delta k_{\max}}^{N}$;
or by the string \texttt{`justFinalPoint'}, which runs those lines
only once, at $k=N=\frac{t_{{\rm specific}}}{\Delta t}$. All calculations
for which the performance%
{} was reported in figure \ref{fig:cpuTimeAndGPUtimeAsFunctionOfT_specific}
were run with \texttt{allPointsOrJustFinalPoint~=~`justFinalPoint'. }Figure
\ref{fig:cpuTimeAndGPUtimeAsFunctionOfT_specific} not only demonstrates
that the benefit of running the main propagation steps on the GPU
rather than the CPU grow with the amount of time for which the system
is being simulated, but also demonstrates very clearly the linear
scaling of the markovian %
tensor propagator algorithm of Makri and Makarov.%
{} 

\section{Discussion\label{sec:discussion}}

It is worth noting that in the program described in this paper, based
on the profiler output in the bottom panel of figure \ref{fig:profilerForCPUvsGPU},
the main propagation calcualtions and each summation used to determine
the numerical values of $\rho$ at a particular time $t$, take a
very small (almost negligible) amount of time to compute. This means
that other parts of the program, which set up the numerical Feynman
integral calculation (such as the calculation of the $\eta$ coefficients
presented in equations \ref{eq:etaKK'generalTrotter} to \ref{eq:etaNkgeneralStrang})
can play an equally important, or dominant role to the overall runtime
of the program (especially if not implemented efficiently). In figure
1 of \cite{Dattani2012b}, it was shown that the calculation of $\eta_{kk^{\prime}}\,,\, k\ne k^{\prime}$
(equation \ref{eq:etaKK'generalTrotter} of this paper), for the same
spectral distribution function used in the calculations presented
in this paper (given by equations \ref{eq:superOhmicJwithGaussianCutoff},
\ref{eq:AforJ(w)usedInThisPaper(fromRamsayGoddenEtc2010prl)} and
\ref{eq:wcForJ(w)usedInThisPaper(fromRamsayGoddenEtc2010prl)}), took
more than 300 seconds to calculate on the CPU used when using Mathematica
8's \texttt{NIntegrate} function to numerically integrate the most
frequently used expressions for these quantities, while it took less
than half a second to compute on the same CPU using Mathematica 8's
\texttt{Sum} function to evaluate the expression introduced in \cite{Dattani2012b}
for the same quantities. For this reason, using the analytic expressions
for the $\eta$ coefficients first given in \cite{Dattani2012b} will
help to maximize the benefit from the GPU version of the program described
in this paper - and if one were to numerically compute the historically
more traditional expressions for these coefficients, the time taken
to execute this numerical integration would in many cases dominate
the execution time for the rest of the program's calculations.

It is also important to discuss the computational overhead associated
with transfering data from the CPU to the GPU and back, especially
the GPU deals with a very large%
\footnote{When the accuracy of the calculations is important, the size of the
data involved in the PMC will ideally be as large as the GPU's memory
will allow.%
} amount of data for the calculations performed by the program described
in this paper. The transferring of the relevant data from the CPU
to the GPU is done by lines 160-163 of the GPU version of the code,
and the transferring of  the result $\{\rho(t_{k})\}$ stored in the
array \texttt{rho}, is transferred back from the GPU to the CPU on
line 171 of that code. The result of the profiler's estimates if runtime
displayed in the bottom panel of figure \ref{fig:profilerForCPUvsGPU}
suggest that the transferring of the data to and from the GPU is performed
in real time. The profiler is perhaps underestimating the runtimes
of at least one of the lines of code performed on the GPU (between
lines 160 and 172). This can be seen by observing that figure \ref{fig:cpuTimeAndGPUtimeAsFunctionOfT_specific}
shows that lines 160 to 172 of the GPU version of the code can take
up to one minute, and the sum of all the runtimes estimated by the
profiler for these lines never reached one minute when these calculations
were redone with the profiler on%
\footnote{Except for the calculations for which the profiler results were displayed
in figure \ref{fig:profilerForCPUvsGPU}, all calculations reported
in this paper were performed with without the profiler running, in
order to remove the confusion that would be introduced due to the
computational overhead of using the profiler (although this overhead
was in fact found to be very small). %
}%
. If it is true that the profiler has underestimated the runtimes
of lines 160-163 in figure \ref{fig:profilerForCPUvsGPU}, which transfer
the relevant data from the CPU to the GPU, this might at least partially
explain why line 165 was reported to have taken significantly longer
than line 166 in the bottom panel of figure \ref{fig:profilerForCPUvsGPU},
despite the top panel of the same figure showing that in the analogous
lines for the CPU version of the code (lines 171 and 172), line 172
is actually faster than 171 ! In any case though, the transfer of
the data to and from the GPU cannot be significant to the overall
runtime of the program, since the exact same data is transfered in
every instance represented by a red circle in figure \ref{fig:cpuTimeAndGPUtimeAsFunctionOfT_specific},
and therefore the overall runtime associated with transferring the
data should be within the scale of the first red circle. The linear
increase seen in the red curve is due to the fact that lines 165 and
166 are repeated more and more times in the \texttt{for} loop as $N$
is increased. 

\medskip{}

The program described in this paper is open source, and users are
encouraged to contribute on Github by implementing techniques to improve
its efficiency, such as the OFPF technique \cite{2001Sim} and the
interpolation method \cite{Dattani2012} mentioned in the introduction,
and the extension to allow for models where the OQS can couple to
more than one QHO of a given frequency using the more general influence
functional first presented in \cite{Nalbach2010b}.

\section{Acknowledgements}

I would like to express my most warm thanks to Professor Mike Giles,
Dr. Wes Armour, Dr. Andrew Richards, and the rest of the \textbf{O}xford
\textbf{S}upercomputing \textbf{C}entre (\textbf{OSC}) and \textbf{O}xford
\textbf{e}-\textbf{R}esearch \textbf{C}entre (\textbf{OeRC}) for allowing
me to use their Viglen/NVIDIA hybrid GPGPU Cluster SKYNET. I am also
delighted to thank Dr. Mihai Duta, Albert Solernou, Steven Young,
and the rest of the Oxford e-Research Centre staff for helping me
get started with using the GPUs on SKYNET, and for being such friendly
and helpful support staff for the OSC and OeRC supercomputers. I also
thank the many members of the MATLAB Central Newsgroup who helped
me develop the MATLAB program described in this paper, particularly
Matt J and James Tursa for introducing me to BSXFUN; and I thank Gallagher
Pryor for his blog entry which immediately brought the power of implementing
BSXFUN on GPUs to my very keen attention. Finally, I thank the Clarendon
Fund and the NSERC/CRSNG of/du Canada for financial support.

\end{document}